\documentclass[11pt,preprint,aps,prb]{revtex4}
\usepackage[dvips]{graphicx}

\newcommand{\modm}{\;(\text{mod}\;m)}

\begin{document}

\title{The solution to the challenge in ``Time-Reversible Random Number Generators'' by Wm.~G.~Hoover and Carol G.~Hoover}

\author{
Federico Ricci-Tersenghi\\
Dipartimento di Fisica, INFN -- Sezione di Roma 1, and CNR-IPCF, UOS di Roma\\
Universit\`a La Sapienza, Piazzale Aldo Moro 5, I-00185 Roma, Italy
}

\date{\today}

\begin{abstract}
I provide the algorithm that solves the challenge proposed by Wm.~G.~Hoover and Carol G.~Hoover in their recent paper ``Time-Reversible Random Number Generators'', arXiv:1305.0961v1, with an explanation on how to derive it analytically.
\end{abstract}

\maketitle

Recently in Ref.~\onlinecite{Hoover} Bill and Carol Hoover have considered a linear congruential generator (LCG) of pseudo-random numbers and have posed the interesting question whether its time-reversed generator can be constructed. Here I provide an explicit construction for a such a time-reversed algorithm.

The general class of LCG that I am considering is defined by the following recursive equations
\begin{eqnarray*}
x_{n+1} &=& a\,x_n + b \modm\;,\\
y_{n+1} &=& a\,y_n + f(x_n) \modm\;.
\end{eqnarray*}
The corresponding pseudo-random number can then be obtained as $R_n = (x_n+m\,y_n)/m^2$.
With a correct choice of the parameters $a$, $b$, $m$ and of the function $f$ the LCG has maximum period (equal to $m^2$) and generates very good pseudo-random numbers.

The LCG considered in Ref.~\onlinecite{Hoover} corresponds to parameters
\[
a=1029\;, \qquad b=1731\;, \qquad m=2048\;,
\]
and to the function
\[
f(x) = 1536\,x + \frac{a\,x+b - (a\,x+b) \modm}{m}\;.
\]
It is worth noticing that for the above parameters choice, $x$ and $y$ are non-negative integer numbers of 11 bits, while $(ax+b)$ is a 22 bits integer number: as a consequence, the second term in $f(x)$ --- the fraction --- returns the 11 most significant bits of $(ax+b)$ shifted down to the position of the 11 less significant bits.

In order to compute the time-reversed of the above LCG, we just need to invert the simplest LCG
\[
x_{n+1} = a\,x_n + b \modm\;.
\]
As long as the LCG has maximum period the inverse LCG certainly exist and can be written as
\[
x_n = c\,x_{n+1} + d \modm\;.
\]
The parameters $c$ and $d$ of the time-reversed LCG must satisfy the following equations
\[
a\,c = 1 \modm\;, \qquad 
c\,b+d = 0 \modm\;, \qquad
a\,d+b = 0 \modm\;.
\]
Even if these looks as three equations in two unknowns, the solution to the first two equations also satisfies the third one: indeed by replacing the solution
\[
d = - c\,b \modm\;,
\]
in the third equation one gets
\[
- a\,c\,b+b = 0 \modm\;,
\]
thanks to the first equation.
For the LCG considered in Ref.~\onlinecite{Hoover} the parameters of the time-reversed LCG are
\[
c = 205 \qquad d = 1497
\]

Reconsidering the general LCG defined by the two coupled maps in $x$ and $y$ written at the beginning, the reversed $x_n$ sequence can be obtained by
\[
x_n = c\,x_{n+1} + d \modm\;.
\]
The map that generates the sequence $y_n$ looks more complicated, but it is actually as the simplest LCG with the parameter $b$ being not constant, but rather depending on the value of $x_n$ (which is already known!). So its time-reversed can be obtained as
\[
y_n = c\,(y_{n+1} - f(x_n)) \modm = c\,(y_{n+1} + m^2 - f(x_n)) \modm\;,
\]
where the last expression is equivalent to the second one, but the $m^2$ term ensures the value in parenthesis is non-negative and so is the full expression to which we have to apply the modulus operation (this is essential at the moment of writing a computer code implementing the above algorithm, since the application of the modulus operation to a negative number is not always well defined).

The following Fortran code first generates the {\em forward} sequence with the {\tt rund} algorithm of Ref.~\onlinecite{Hoover}, then generates the {\em backward} sequence with the reversed algorithms of {\tt rund} and finally checks that the latter sequence is indeed the reversed of the former.

\begin{verbatim}
program reverseRund
implicit none
integer, parameter      :: imax = 4194304
integer,dimension(imax) :: forwx, forwy, backx, backy
integer                 :: i, j, intx, inty, oldx, n

intx = 0
inty = 0
do n=1,imax
   i = 1029*intx + 1731
   j = i + 1029*inty + 507*intx - 1731
   intx = mod(i,2048)
   j = j + (i - intx)/2048
   inty = mod(j,2048)
   forwx(n) = intx
   forwy(n) = inty
end do

intx = 0
inty = 0
do n=1,imax
   oldx = mod(205 * intx + 1497, 2048)
   inty = inty + imax - 1536 * oldx - (1029 * oldx + 1731 - intx) / 2048
   inty = mod(205 * inty, 2048)
   intx = oldx
   backx(n) = intx
   backy(n) = inty
end do

do n=1,imax-1
   if (backx(n) /= forwx(imax-n) .or. backy(n) /= forwy(imax-n)) then
     print *,'error'
     exit
   end if
end do
end program reverseRund
\end{verbatim}

\end{document}